\documentclass{emulateapj}
\usepackage{bm}
\usepackage{color}
\usepackage{graphicx}
\usepackage{natbib}
\begin{document}
\title{The isotropy problem of Sub-ankle Ultra-high energy cosmic rays}
\author{Rahul Kumar}
\author{David Eichler}
\affiliation{Physics Department, Ben-Gurion University, Be'er-Sheba 84105, Israel}

\begin{abstract}
We study the time dependent propagation of sub-ankle ultra-high energy cosmic rays (UHECRs) originating from point-like Galactic sources. We show that drift in the Galactic magnetic field (GMF) may play an important role in the propagation of UHECRs and their measured anisotropy, particularly when the transport is anisotropic. To fully account for the discreteness of UHECR sources in space and time, a Monte Carlo method is used to randomly place sources in the Galaxy. The low anisotropy measured by Auger is not generally characteristic of the theoretical models, given that the sources are distributed in proportion to the star formation rate, but it can possibly be understood as a) intermittency effects due to the discrete nature of the sources or, with extreme parameters, b) a cancellation of drift current along a current sheet with the outward radial diffusive flux. We conclude that it is possible to interpret the Galactic sub-ankle CR flux as being due entirely to intermittent discrete Galactic sources distributed in proportion to star formation, but only with probability of roughly 35 percent, of which the spectrum is in accord with observations about 30 percent of the time.

An alternative explanation for the low anisotropy may be that they are mostly extragalactic and/or heavy.
\end{abstract}
\section{Introduction}
It is widely suspected that the abrupt flattening of the cosmic ray (CR) spectrum above 4 EeV, the so-called ankle \citep{auger_spectrum} of the CR spectrum, is due to transition from the Galactic to extragalactic origin of CRs. Either way, the origin of EeV UHECRs is still an open question. Blast waves from supernova remnants are believed to be unable to accelerate CR protons to energy beyond $10^{14.5}$ eV \citep{Lagage}, though Jokipii (private communication) has argued that higher energies may be attainable on equatorial drift trajectories if the magnetic field is sufficiently ordered. Special subsets of supernovae may be powerful enough that EeV CR energies might be attainable, even for protons, such as GRB \citep{levinson1993,Eichler2011}. Supernova explosions of massive stars with strong stellar wind \citep{Biermann}  can possibly be the sources for CRs up to the ankle if they are iron-like, though the AUGER collaboration reports light composition in this energy range. In this case, the source rate might be much less than for supernovae as a whole, and, as shown below, the source rate plays a role in determining the anisotropy and spectrum at a typical moment.  In this paper, we assume a Galactic origin of CRs up to the ankle for various source rates, ranging from the total supernova rate of $\sim 10^{-2} yr^{-1}$ to $10^{-6} yr^{-1}$.

Recent observations by the \cite{auger2011} suggest a nearly isotropic flux of sub-ankle UHECRs. Specifically, the observed large scale anisotropy in the arrival direction of UHECRs is less than 2.3$\%$ \footnote{the statistical significance has undergone some reevaluation, but this is the number we adopt in this paper} in 2-4 EeV energy band. The anisotropy measurements seem to challenge the notion that the sub-ankle UHECRs are mostly Galactic, if their sources are distributed in proportion to the star formation rate in the Galaxy. In a previous paper, \cite{paper1} showed that the anisotropy predicted by an isotropic diffusion model due to discrete sources in the Galaxy is on average much higher than the measured value. The premises of their model led the authors to conclude that a) the mean free path of the UHECRs is small compared to the proton Larmor radius, or b) intermittency is an important factor and we are living in a rare lull in UHECR production, or c) majority of sub ankle CRs have extragalactic origin. Estimates of anisotropy in several other models \cite[e.g.]{calvez,Giaciniti} are also higher than the observed limit for light sub-ankle primaries. We refer to this discrepancy between measured value and theoretical models as the UHECR isotropy problem.

   The purpose of this paper is to reevaluate the above claims taking into consideration a) anisotropic diffusion and b) drift, both of which we show to be important. We make the working assumption that the UHECRs in the 2-4 EeV energy band are protons and largely due to Galactic sources.\footnote{At $10^{18}$ eV the composition of UHECRs is light, though not necessarily dominated by protons \cite{auger_composition, hires_composition}. A heavier composition of average nuclear charge $Z$ at energy $E_{Z}$ would behave like a proton dominated UHECRs of energy $E = E_{Z}/Z$, and would imply a lower anisotropy than a proton dominated UHECRs at the same energy.} Once injected by the sources, propagation of CRs from their sources to the Earth would depend merely on the interaction of CRs with the interstellar medium regardless of the nature of the sources.

    We study the propagation of UHECRs in a partially ordered inhomogeneous GMF. The gyration of UHECRs in the GMF reduces their transport rate across the magnetic field lines, and makes UHECRs propagation in the Galaxy anisotropic \citep{evoli2012}. In addition, an inhomogeneous GMF causes UHECRs to drift and their propagation, for a mean free path that is comparable or larger than Larmor radius, can not adequately be described by diffusion alone. In order to take anisotropic diffusion and drift into account we study the propagation of UHECRs in various configurations of the GMF in a test particle simulation. We show that the drift can significantly alter UHECR propagation and the measured anisotropy in a counterintuitive way.

\section{propagation of UHECR\lowercase{s} in the Galaxy}
 The turbulent GMF causes UHECRs to scatter and the propagation of CRs in the GMF can phenomenologically be described by diffusion. The regular component of the GMF (directed along the spiral arms and here assumed to be toroidal) breaks isotropy of the diffusion of CRs with respect to the source since the cross field transport of CRs is reduced due to gyration. Assuming diffusive propagation of UHECRs, the differential density N of UHECRs that are released instantaneously by a point-like source at $\vec{r}=\vec{r_s}$ is governed by

\begin{equation}
   \frac{\partial N}{\partial t}=\nabla D \nabla N +Q(E)\delta(t)\delta(\vec{r}-\vec{r}_s),
\end{equation}
 where D is diffusion tensor and $Q(E)$ is differential production rate of UHECRs at energy E. We have ignored the energy loss due to photo-secondary production and spallation, since the time scale ($\gtrsim 10^8$ years) for both are much larger than the escape time ($\lesssim 10^6$ years) of UHECRs at the sub-ankle energies. We assume the classical scattering limit for diffusion of charged particles in a magnetic field, that is to say CRs move in the regular GMF between subsequent instantaneous isotropic scattering events. In this limit, the diffusion coefficients parallel and perpendicular to the magnetic field lines at any given energy are $D_{\parallel}=c\lambda_{mfp}/3$ and $D_{\perp}=c\lambda_{mfp}/[3(1+\lambda_{mfp}^2/r_L^2)]$ respectively \citep{Axford}, where $r_L$ is Larmor radius, $c$ is the speed of light, and $\lambda_{mfp}$ is the mean distance travelled by UHECRs between two subsequent random scattering events. In galacto-centric polar coordinates, where z-axis is perpendicular to the Galactic plane, the diffusion coefficients can be written as: $D_{\rho\rho} = D_{zz}=D_\perp$, $D_{\phi\phi}=D_\parallel$, and off-diagonal terms $D_{\rho z}=-D_{z\rho}=cr_L(\lambda^2_{mfp}/r_L^2)/[3(1+\lambda_{mfp}^2/r_L^2)]$, which are known as drift coefficients. If the mean free path is $ \gtrsim r_L$, drift coefficients become comparable to the axial diffusion coefficients and must be taken into consideration in the propagation models.

\subsection{Test Particle Simulation}
 We simulate the diffusive propagation of the UHECRs in the classical scattering limit, as described above, with a large number of test particles. We use leapfrog integration method to compute trajectories of the test particles in a prescribed GMF. Particles are moved in computed trajectories between two scattering events, in which the directions of their velocity vectors are randomised. We assume a toroidal GMF
    \begin{equation}
       \mathbf{B} =B_{\phi}(\rho,z) \hat{\phi}=B_{0}B_{\phi}^{\rho}(\rho)B_{\phi}^{z}(z) \hat{\phi}
    \end{equation}
    which is assumed to weaken with distance from the Galactic plane and the Galactic center. As suggested by various models of the GMF, gradient of the GMF in the radial direction is assumed to be rather small as compared to the gradient along the z-axis. Specifically, we choose
  \begin{equation}
    B_{\phi}(\rho)=\exp(-(\max(\rho,\rho_{c})-\rho_{\odot})/\rho_{0})
  \end{equation}
      where $\rho_{\odot}$ is distance of the Sun $\approx$ 8.5 kpc, $B_0=10 \; \mu G $,  and the parameters $\rho_{c}$ and $\rho_{0}$ are taken to be 3 kpc and 10 kpc respectively \citep{GMF}. Models of the GMFs generally suggest that the GMF decreases exponentially with $z$, with a scale height varying from 1 kpc to few kpcs \citep{Farrar_GMF}. However, if the GMF is stretched out by a Galactic wind \citep{Everett}, it is quite reasonable to believe that the GMF decreases rather slowly with distance from the Galactic plane, supposedly as $1/z$. The variation of the GMF with z can have large uncertainties which would be reflected on the confinement and drift of UHECRs in the Galaxy. We consider few possibilities for the z-dependence of the $B_\phi(z)$, that represent a weak or strong halo GMF. In particular, we consider five different $B^z_\phi(z)$:   MF1: $exp(-|z|/ 1.5 kpc)$, MF2: $exp(-|z|/3 kpc)$, MF3: $1/(1+|z|/1.5 kpc)$, MF4: $1/(1+|z|/3 kpc)$, and MF5: $1/(1+|z|/10 kpc)$.

      As shown later, significant contribution to the CR flux comes from sources that are within few kilo-parsecs and hence the details of the GMF near the Galactic center, such as Fermi bubble \cite[e.g.][]{fermi_bubble}, would not affect our results and have not been considered in the present analysis.

   The mean free path is assumed to scale with local strength of regular component of the GMF. The scaling implies that UHECR diffusion rate is larger in the halo as compared to the disk. Particles reaching few kpc into the halo, where the GMF is no longer strong enough to curve their trajectories back towards the Galactic plane, escape out of the Galaxy. It implies that the UHECRs that originated few kpc away preferentially escape out of the galaxy before reaching us, hence limiting our UHECR source visibility in the Galaxy.

\subsection{Anisotropy}
The CR intensity $\Phi(\mathbf{n})$ in any direction $\mathbf{n}$ can be written as
 \begin{equation}
    \Phi(\mathbf{n})=\Phi_0+\Phi_1 \mathbf{n}\cdot \mathbf{d},
 \end{equation}
where $\mathbf{d}$ is the dipole unit vector. The dipole anisotropy $\delta$ is defined as maximal intensity contrast:
  \begin{equation}
      \delta=\frac{\Phi_{max}-\Phi_{min}}{\Phi_{max}+\Phi_{min}}=\frac{\Phi_1}{\Phi_0}
  \end{equation}
  The Auger collaboration reports an upper limit on $\delta$ in various UHECR energy bands.

  In our simulation we compute mean velocity $\bar{\mathbf{u}}$ (in units of c ) averaged over all particles in the vicinity of the Earth that lie within a distance $\lesssim r_L$. For a given $\Phi(n)$,
   \begin{equation}
       \bar{\mathbf{u}}=\int_{-1}^{1} \Phi_1 \cos^2\theta d \cos \theta / \int_{-1}^{1} \Phi_0 d \cos \theta = \Phi_1/3\Phi_0
   \end{equation}

 where $\cos \theta=\mathbf{n}\cdot \mathbf{d}$. The maximal intensity contrast or the dipole anisotropy $\delta$ ( hereafter referred to as anisotropy) inferred from the simulation is 3$\bar{\mathbf{u}}$, which is then compared against the measured anisotropy by Auger.

 In order to compute $\bar{\mathbf{u}}$ down to about 1$\%$ we start off with a large number of particles ($\gtrsim 10^6$) that are released isotropically from the source at once. We further split our test particles into particles of lower weights as they happen to be nearer to the Galactic solar circle, hence generating enough statistics to compute $\bar{\mathbf{u}}$ down to 1$\%$.

\subsection{Source distribution}
It is plausible to believe that if the sub-ankle UHECRs are produced in compact Galactic sources, such as long GRBs, then UHECR sources are distributed in proportion to the star formation rate in the Galaxy. In particular, spatial probability distribution in galactocentric radius,
\begin{equation}
  P(\rho_{GC})=\frac{2\rho_{GC}}{\rho_0^2}\exp\left(-\frac{\rho_{GC}^2}{\rho_0^2}\right)
  \label{distribution}
\end{equation}
    with scale $\rho_0=5$ kpc is adopted as an approximation to the distribution of baryons and star formation in the Galaxy, and connotatively, approximation to the distribution of UHECR sources in the Galaxy.

\subsection{Drift in the Galactic magnetic field}
It is well known that a significant variation in magnetic field over a distance comparable to Larmor radius of a charged particle leads to drift motion of its guiding center. The drift velocity of protons of energy E in a magnetic field $\mathbf{B}$, averaged over a density distribution function to the lowest order is $E(\nabla \times \mathbf{B}/B^2)/3eB^2$, which is sum of well known gradient and curvature drifts \citep{drift_burger}.

The gradient of GMF along the z-axis causes UHECRs to drift in the radial direction. Since the $\nabla_z B$ changes sign across the Galactic plane, the radial drift due to $\nabla_z B$ is oppositely directed in the upper ($z>0$) and lower ($z<0$) Galactic halos. Specifically, UHECRs drift towards the Galactic center for $z>0$ and away from it for $z<0$. This effect of  drift would create a density gradient at the Earth's location in the positive (negative) z-direction if the UHECR source lie within (outside) the Galactic solar circle. Locally, a density gradient creates a net current due to gyration of charged particles in local magnetic field, known as diamagnetic current. The diamagnetic current can be approximated by $E(\mathbf{B} \times \nabla N)/3qNB^2$ which is correct up to the first order in anisotropy in density distribution. Taking the diamagnetic current into consideration, a gradient along the z-axis implies that there is a net radial current of UHECRs towards their sources which subdues the diffusive flux pointing away from the sources. In addition, radial UHECR density gradient implies a diamagnetic current along the z-axis which adds to the diffusive flux along the z-axis due to the density gradient created by radial drift. This additional anisotropy along the z-axis originates as a result of drift and has not been considered in the previous studies.

 In addition to the drift due to gradient of the GMF along the z-axis, gradient of the GMF in the radial direction also contributes to the UHECR drift. The strength of gradient drift is inversely proportional to the scale height of the GMF, so the contribution to the drift due to radial gradient is comparatively smaller than the drift due to gradient along the z-axis in most cases considered here. Also, drift due to the curvature of the GMF is locally weaker compared to the $\nabla_z B$ drift. The curvature and radial gradient drift affects the propagation in rather longer time scale, at least within few kpc of the Earth, by slowly drifting UHECRs along the z-axis.

\subsection{Infinite source rate limit}
\begin{figure}
\begin{center}
\includegraphics[scale=0.5]{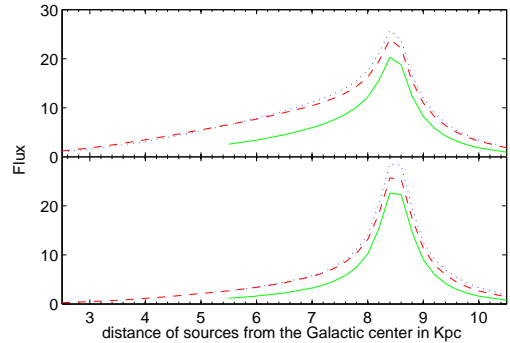}
\end{center}
\caption{UHECRs flux (in arbitrary units) due to all sources that lie at a distance $d$ from the Galactic center is plotted against $d$, assuming that the sources are distributed in proportion to the star formation rate (eq \ref{distribution}). In the top panel $\lambda_{mfp}=r_L$ and in the bottom panel $\lambda_{mfp}=3r_L$. Solid (green), Dashed (red), and dotted (blue) curves are for the GMF configurations MF1, MF2, and MF3 respectively in each panel. Here, and henceforth, energy is 2.4 EeV. Drift as well as faster diffusion in the halo leads to faster escape of UHECRs, extenuating the contribution of distant sources to the observed UHECRs flux.}
\label{flux}
\end{figure}

We compute the time average value of UHECR flux anisotropy in the case of continuum production of Galactic UHECRs in space and time, assuming that the sources are distributed in proportion to the star formation rate approximated by eq \ref{distribution}. In figure \ref{flux}, we have plotted relative contribution to the local UHECR flux by all sources that are at a distance $d$ from the Galactic center for three different configurations of the GMF. It serves to show that most of the local UHECRs originate in sources within few kilo-parsecs from the Earth, even for a strong halo magnetic field such as MF3. UHECRs that originate in sources beyond few kilo-parsecs preferentially escape out of the face of the Galaxy before reaching us, due to faster diffusion in weaker halo GMF and the drift. A relatively stronger halo GMF, such as MF2 and MF3, as compared to MF1, slightly enhances the confinement of UHECRs and hence increasing the mean distance they travel along the galactic plane before escaping out. In the case of a larger mean free path, such as $\lambda_{mfp}=3r_L$ (bottom panel), the radial transport rate is reduced and the escape due to drift is enhanced which reduces our UHECRs source visibility even further as compared to the Bohmian diffusion ($ \lambda_{mfp}=r_L$).

\begin{figure}
\begin{center}
\includegraphics[scale=0.5]{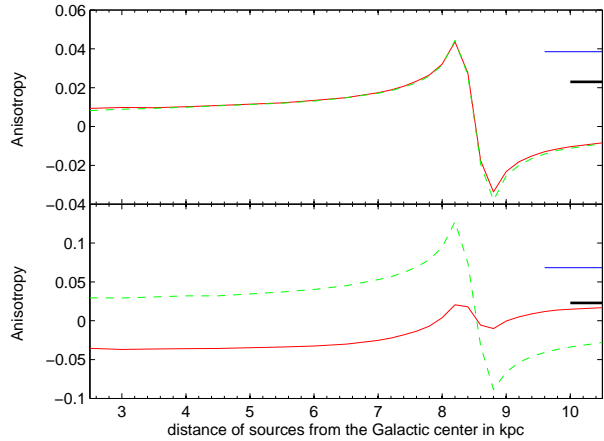}
\end{center}
\caption{Anisotropy (in arbitrary units) due to all sources that lie at a distance $d$ from the Galactic center is plotted against $d$. Here the GMF is assumed to be MF3. In the top panel $\lambda_{mfp}=r_L$ and in the bottom panel $\lambda_{mfp}=3r_L$. Solid (red) and Dashed (green) curves represent anisotropy along the radial and z-direction respectively. The thick black bar in the right indicates observed 99$\%$ upper limit set by Auger and the thin blue line indicates the mean anisotropy assuming that the sources are distributed in proportion to the star formation rate (eq \ref{distribution}).}
\label{anis}
\end{figure}

 To illustrate the effect of drift on the anisotropy, in figure \ref{anis} we have plotted radial and z-component of the local anisotropy due to the sources at a distance $d$ from the Galactic center, for the GMF configuration MF3. In the top panel, where diffusion is assumed to be Bohmian, we see that the anisotropy has a comparable component along the z-axis, in addition to the radial anisotropy due to diamagnetic current along the z-axis. As the mean free path becomes larger, propagation of UHECRs becomes more anisotropic and the drift contributes more significantly to the anisotropy. Specifically, for $\lambda_{mfp}=3r_L$, as shown in the bottom panel, radial diamagnetic current becomes significantly larger than the radial diffusive flux and the net radial anisotropy reverses its direction to point towards the source for the sources that lie within the Galactic solar circle. The radial diamagnetic current becomes large due to large local density gradient along the z-axis which also implies a large diffusive flux along the z-axis. This enhances the anisotropy along the z-axis, hence raising the total anisotropy considerably.

In table 1, we have listed the mean value of the anisotropy for five different configurations of the GMF and two different values of the mean free path. As is evident from the table \ref{table1}, the anisotropy goes down slightly for a stronger halo GMF due to a relatively better confinement that enables us to see sources further in the Galaxy. As the mean free path is increased the total anisotropy goes up due to the drift. For all choices of the GMFs and mean free path considered here the total anisotropy at 2.4 EeV remains significantly higher than the 99$\%$ limit set by Auger.

\begin{table}
\begin{center}
\begin{tabular}{| l | l | l | l | l | l | l |}
  \hline
       $B_\phi^z(z)$ & \multicolumn{3}{|c|}{$\lambda_{mfp}=r_L$} & \multicolumn{3}{|c|}{$\lambda_{mfp}=3r_L$}  \\
  \hline
	     & $\delta_\rho$ & $\delta_z$ & $\delta_{tot}$& $\delta_\rho$ & $\delta_z$ & $\delta_{tot}$ \\
  \hline
       $\exp(-|z| / 1.5 kpc)$  & 3.4$\%$ & 3.1$\%$ & 4.6$\%$ & 0.0 $\%$ & 7.4 $\%$ & 7.4 $\%$ \\
  \hline
       $\exp (-|z|/3 kpc )$  & 3.1$\%$ & 2.6$\%$ & 4.0$\%$ & 1.7$\%$ & 6.5 $\%$& 6.7 $\%$\\
  \hline
       $1/(1+|z|/1.5 kpc)$  & 2.9$\%$ & 2.6$\%$ & 3.9$\%$ & -0.4$\%$ & 6.8 $\%$& 6.8 $\%$\\
  \hline
       $1/(1+|z|/3 kpc)$  & 2.8$\%$ & 2.2$\%$ & 3.6$\%$ & 1.5$\%$ & 5.9 $\%$& 6.0$\%$\\
  \hline
       $1/(1+|z|/10 kpc)$  & 2.6$\%$ & 1.8$\%$ & 3.2$\%$ & 3.1$\%$ & 4.8 $\%$& 5.7$\%$\\
  \hline
\end{tabular}
\caption{ Anisotropy in the arrival direction of UHECRs protons of 2.4 EeV in the infinite source rate limit for GMFs of differing strength in the halo, assuming the source distribution given by eq \ref{distribution}. $\delta_\rho$ and $\delta_z$ are anisotropies in the radial and z-direction respectively. $\delta_{tot}=\sqrt{\delta_\rho^2+\delta_z^2}$}.
\end{center}
\label{table1}
\end{table}

\begin{table}
\begin{center}
\begin{tabular}{| l | l | l | l | l | l | l |}
  \hline
       $|B_\phi^z(z)|$ & \multicolumn{2}{|c|}{$\lambda_{mfp}=r_L$} & \multicolumn{2}{|c|}{$\lambda_{mfp}=3r_L$}  \\
  \hline
	     & $\delta^{out}_\rho$ & $\delta^{in}_\rho$ & $\delta^{out}_\rho$ & $\delta^{in}_\rho$  \\
  \hline
       $\exp(-|z|/ 1.5 kpc)$  & 5.4$\%$ & 1.1$\%$ & 19.1$\%$ & -21.2 $\%$  \\
  \hline
       $\exp (-|z|/3 kpc )$  & 3.9$\%$ & 1.9$\%$ & 12.2$\%$ & -13.5$\%$ \\
  \hline
       $1/(1+|z|/1.5 kpc)$  & 2.7$\%$ & 2.9$\%$ & 12.1$\%$ & -16.7$\%$\\
  \hline
       $1/(1+|z|/3 kpc)$  & 3.1$\%$ & 2.5$\%$ & 9.7$\%$ & -11.7$\%$\\
  \hline

\end{tabular}
\caption{ Anisotropy in the infinite source rate limit as in the table 1. Here The GMF is assumed to reverse its direction across the Galactic plane. $\delta^{out}_\rho$ and $\delta^{in}_\rho$ are the radial anisotropy for the cases where $\nabla \times \mathbf{B}$ or the current is along $\hat{\mathbf{\rho}}$ and $-\hat{\mathbf{\rho}}$ respectively.}
\end{center}
\label{table_current_sheet}
\end{table}

We consider the possibility of a Galactic current sheet or GMF changing sign across the Galactic plane \citep{Zweibel}. In this scenario, anisotropy along the z-axis vanishes due to symmetry of the propagation about the Galactic plane and drift of UHECRs along the current sheet contributes to the net radial anisotropy. Radial anisotropy for the GMFs of different halo strengths and direction of Galactic current is listed in the table \ref{table_current_sheet}. An inward drift can partially cancel the diffusive flux and hence reducing the net radial anisotropy that can possibly be below the measured value, which is the case with the Bohmian diffusion and the GMFs  MF1 and MF2. On the other hand, an outward drift adds to the diffusive flux and hence raises the net anisotropy further. If the mean free path is considerably larger than the Larmor radius, drift becomes dominant and UHECRs stream along the current sheet that leads to a large anisotropy well above the observed limit.

\subsection{Discrete sources}
   If we assume that the UHECR production takes place in spatially localised events in the Galaxy, such as relativistic shocks in GRBs, that lasts for few thousands years, discreteness of UHECR sources in space and time would imply that anisotropy and flux at any instant can deviate significantly from their time average value described in the previous section.

   In order to take into account discreteness of the sources in space and time we first compute anisotropy at time intervals of 10 years due to point-like sources that are placed at finely spaced discrete locations (source separation is taken to be 0.2 kpc up to 3 kpc from the Earth and 0.5 kpc otherwise, and each source is assumed to be active for 10 thousand years). We use Monte Carlo method to randomly place sources in the Galaxy at these discrete locations with the probability distribution given by eq \ref{distribution} for a given source rate. We sum individual contributions of each sources to flux at the Earth to compute the net flux. Similarly, Anisotropies is computed by vectorially summing the anisotropies from individual sources weighted over their flux.

\begin{figure}
\begin{center}
\includegraphics[scale=0.5]{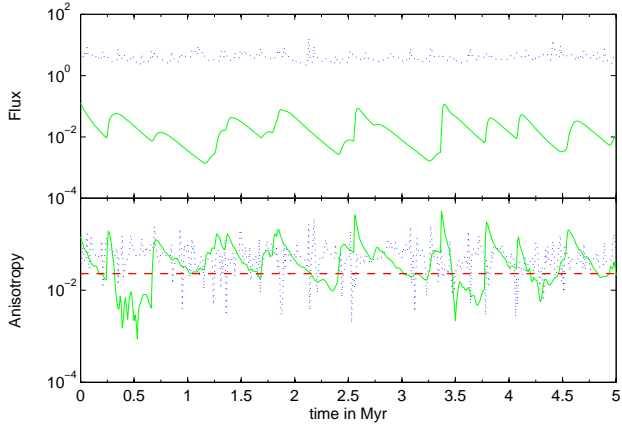}
\end{center}
\caption{Top panel: 5 Myr light curve of the local proton flux in arbitrary units for two different source rates: the dotted(blue) curve is for source rate $10^{-2} yr^{-1}$, and the solid (green) curve is for $10^{-4} yr^{-1}$.  Bottom panel: Temporal variation of the anisotropy is plotted for two different source rates as in the top panel. The dashed (red) line indicates 99$\%$ upper limit set by Auger. Here the GMF is chosen to be MF1 and $\lambda_{mfp}=r_L$. }
\label{DS1}
\end{figure}

\begin{figure}
\begin{center}
\includegraphics[scale=0.5]{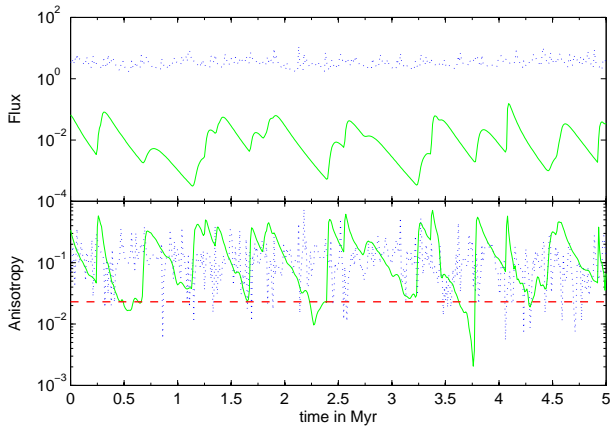}
\end{center}
\caption{ Temporal variation of anisotropy and flux as in the figure \ref{DS1}, but for $\lambda_{mfp}=3r_L$. All other details are same as in the figure \ref{DS1}.}
\label{DS2}
\end{figure}

   In figure \ref{DS1} and \ref{DS2}, we have plotted temporal variation of flux and anisotropy for two different source rates, namely, $10^{-2} yr^{-1}$ and $10^{-4} yr^{-1}$ for the magnetic field configuration MF1. As evident from the flux curves in figure \ref{DS1}, Bohmian diffusion of UHECRs implies that if the UHECR production events in the Galaxy is $10^{-4} yr^{-1}$, we spend significant amount of time in lulls of UHECR production where the UHECR density is much lower than its time average value. The flux observed at the Earth can fluctuate by two orders of magnitude for the source rate $10^{-4} yr^{-1}$. The anisotropy during the lulls also goes down as the UHECR spread out in the Galaxy. In particular, for Bohmian diffusion the anisotropy goes below 2.3$\%$ about 30$\%$ of time for source rate $10^{-4} yr^{-1}$. However, chances of being in one of these lull fades if the source rate is high. The UHECR density in interstellar medium is rejuvenated by newly injected UHECRs from nearby recent events and the anisotropy generally remains high. For $\lambda_{mfp}=3r_L$, anisotropy is generally higher than the Bohmian diffusion for a given source rate and hence the probability to meet the observation is smaller.

\begin{figure}
\begin{center}
\includegraphics[scale=0.5]{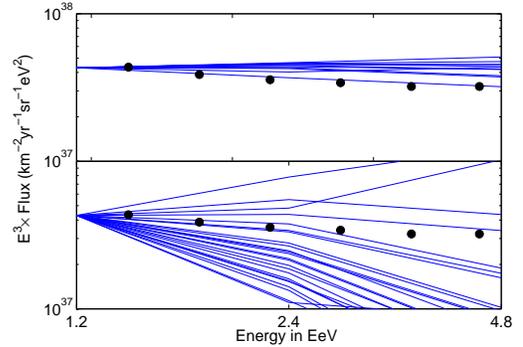}
\end{center}
\caption{ Proton spectra at the solar circle when anisotropy is below 2.3$\%$ at 2.4 EeV. The thin solid curves show spectra for randomly selected realisations of the source distribution. The black dots indicate the flux measured by Auger. In all cases, flux at 1.2 EeV is normalised to $4.3\times10^{37} km^{-2}yr^{-1}sr^{-1}eV^2$. The mean free path $\lambda_{mfp}=r_L$ and the GMF configuration is MF1. The Top and the bottom panels are for the source rate $10^{-2} yr^{-1}$ and $10^{-4} yr^{-1}$ respectively. The source spectrum is assumed to be $E^{-2}$.}
\label{spectrum}
\end{figure}

 The observed anisotropy limit in the 2-4 EeV band can be met intermittently in the lulls of UHECR production for a range of parameters. A lower source rate implies a relatively higher fraction of the lull periods. However, for a very small source rate, for which the lull period extend much longer than the escape time of the EeV UHECRs ($\sim 10^5 years$), spectrum cuts off exponentially at some energy below EeV as the UHECRs escape out of the Galaxy and doesn't meet the observed power law spectrum which is slightly steeper than $E^{-3}$. In figure \ref{spectrum}, we have plotted spectra sources rates $10^{-2}$ and $10^{-4}$ during the period when anisotropy is below 2.3$\%$. As can be seen, for a source rate $\lesssim 10^{-4}$ spectra are generally steeper than observed $E^{-3}$ spectrum in the production lulls when the anisotropy is below the measured upper limit. The spectrum is not significantly steeper than the observed spectra for about 35 $\%$ of the time.

Note however, that if we assume that trans-ankle CR are mostly {\it extragalactic} and that the contribution below the ankle is that expected from extrapolation of the trans-ankle power law spectrum to below the ankle,
the implication is that the Galactic component cuts off sharply just below the ankle, and this would accommodate a source rate that is less than the escape rate of particles near the ankle.

\section{Summary and Conclusions}
We have simulated time-dependent propagation of UHECRs from point-like sources in the Galaxy by a large number of test particles. The simulation includes a) anisotropic diffusion, b) drift, c) spatially varying diffusion, d) source discreteness, and e) temporal variation.

The gyration of UHECRs in the GMF, when not interrupted by scattering, reduces the effective mean free path of UHECRs across the field lines which are mostly toroidal. A smaller perpendicular mean free path might be expected to reduce the anisotropy. In addition, it should increase the confinement of CRs in the Galaxy, hence reducing the fluctuation in flux and anisotropy because of a larger number of sources contributing to the flux at any instant. However, as shown above, as the parallel mean free path becomes comparable or larger than Larmor radius, drift due to inhomogeneity of the GMF significantly alters the propagation. Drift enhances the particle escape and observed anisotropy contrary to what might be intuitively expected merely from consideration of anisotropic diffusion alone. In other words, the off-diagonal terms in the diffusion tensor contribute a component of flux that offsets the effects of differences among the diagonal components.

We have found that if the sources of UHECRs are intermittent, the anisotropy can be well below 2.3$\%$ upper limit during the lulls and can possibly explain the observation - but only about 30 percent of the time, and then only if the source rate R is well below $10^{-2}$ per year.  On the other hand, assuming that we live in an era of intermittent isotropy, (i.e.  unusually low anisotropy) implies that the energy spectrum of the UHECR is not likely to be the time average.  Fitting both the spectrum and the anisotropy limits simultaneously is more difficult than fitting either one separately, and we find that this is most likely to occur when the source rate obeys $R\lesssim 10^{-4}$ per year. In this case, the anisotropy is below 2.3\% about 30 percent of the time, when discreteness anisotropy somewhat offsets systematic radial anisotropy inherent in the source distribution. During these periods of low anisotropy, the spectrum is typically steeper than its time average, but is consistent with the observed spectrum at least 30 percent of the time. In conclusion, it is possible to accommodate the AUGER data with purely Galactic sources some - but not most - of the time, and it becomes a somewhat imponderable question whether this is an attractive option.

  The possibility remains that the  UHECRs in 2-4 EeV are composed mainly of component that originates beyond the solar circle, e.g. an extragalactic component or a "quasi"-extragalactic component -  one that originates well beyond the solar circle - e.g.  at a Galactic wind termination shock (Jokipii and Morfill, 1987).  This is a natural proposition given that those above the ankle seems to represent a separate component.  If we extrapolate the $E^{-2.7}$ power law above the ankle to sub-ankle energies,  it stands to reason that at 2 to 4 EeV, i.e. just below the ankle, most of the UHECRs are also extragalactic, and that only $\sim 30$ percent are Galactic.\footnote{If the trans-ankle component is truly extragalactic, then losses due to pair production again the cosmic microwave background in any case causes a steepening below 4 EeV, further limiting the contribution of a Galactic component}.  In this case, the Galactic component should probably be interpreted as being exponentially cutoff with a cutoff energy at $\lesssim 2 EeV$.   If we further assume that the extragalactic component is light, then we are free to assume whatever we wish about the Galactic component, and the various theories of Galactic sources that would marginally accelerate iron-like nuclei up to 1 Eev become consistent even with the AUGER claim of light composition. Heavy UHECR at EeV energies, of course, can have a much lower anisotropy that protons, so the measured anisotropy, in this scenario, can be extremely low. If AUGER fails to ever detect any anisotropy, while continuing to measure light composition at $1 \le E \le 4$ EeV, it would be entirely consistent with this scenario - that the ankle is merely the termination of a sharply decreasing Galactic component giving way to an extragalactic component that dominates at 4 EeV. The composition and energy dependence of the composition of this extragalactic component would be an entirely separate matter.

We acknowledge support from the Israel-U.S. Binational Science foundation, the Israeli Science Foundation, and the Joan and Robert Arnow Chair of Theoretical Astrophysics.
\bibliography{CR_anis_ref}

\end{document}